\documentclass{goodclass}
\usepackage{amssymb}

\begin{document}

\title{Interactions in Intersecting Brane Models}

\author{A.W.Owen}

\address{Institute for Particle Physics Phenomenology, University of Durham,
  Science Laboratories, South Road, Durham, DH1 3LE\\ 
E-mail: a.w.owen@dur.ac.uk}

%%%%%%%%%%%%%%%%%%%%%%%%%%%%%%%%%%%%%%%%%%%%%%%%%%%%%%%%%%%%%%
% You may repeat \author \address as often as necessary      %
%%%%%%%%%%%%%%%%%%%%%%%%%%%%%%%%%%%%%%%%%%%%%%%%%%%%%%%%%%%%%%

\maketitle

\abstracts{We calculate tree level three and four point scattering
amplitudes in type II string models with matter fields localized at the
intersections of D-brane wrapping cycles. The analysis of the three point
amplitude is performed in the context of Yukawa couplings and it is seen that
a natural mechanism for the generation of a mass hierarchy arises. The four
point amplitude for fermions at the intersection of four independent stacks
of D-branes is then determined.}

\section{Introduction}

The intersecting brane scenario has been remarkably successful in producing
semi-realistic models. For example, models similar to the Standard Model can be
obtained~\cite{ralph1,ibws,chiral,ralphie,just,just2,ckok1,ckok2} and viable
constructions with N=1 supersymmetry have been
developed~\cite{cvetic2,cvetic1,ralphsusy,honecker,cvetic6}. Furthermore,
they also provide a rather attractive topological explanation of family replication.

In this talk, we deal with the computation of the general three point and
four point amplitudes of string states localised at the intersections of
D6-branes wrapping $T^{2} \times T^{2} \times T^{2}$. Our calculations are
based on and extend work presented in~\cite{mirjam,paper2} and may easily 
be adapted to other scenarios involving intersecting branes. A more thorough
discussion of the general four point amplitude and the generalisation to
N-point amplitudes will be presented in~\cite{paper3}, where a detailed list of references is also provided.

\section{The general three point amplitude}\label{sec:3pt} 
We begin by discussing the three point amplitude in the context of Yukawa
interactions. 

\subsection{Yukawa interactions}\label{subsec:yuks}
\begin{figure}
\begin{center}
\epsfig{file=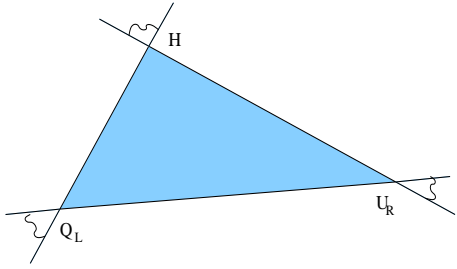, height=35mm, width=60mm} 
\caption{Yukawa interaction. \label{fig:yuk}}  
\end{center}
\end{figure}

Due to our choice of internal space, the amplitude 
factorises into three identical contributions, one from each torus
subfactor. Therefore, concentrating on a single torus, the string states
are localised at the vertices of a triangle whose boundary consists of a
single internal dimension from each of the D6-branes, as depicted in
figure~\ref{fig:yuk}. One would expect the amplitude to be dominated by an
instanton, and thus be proportional to $e^{-\frac{1}{2\pi \alpha}A}$ where $A$
is the area of the triangle. This will be borne out in the following calculation.

 We denote the spacetime coordinates of a single torus
by $X=X^{1}+iX^{2}$ and $ \bar{X}=X^{1}-iX^{2}$.
The bosonic field $X$ can be split up into a classical piece, $X_{cl}$, and a
quantum fluctuation, $X_{qu}$. The amplitude then 
factorises into classical and quantum contributions,
\begin{equation}
Z=\sum_{\langle X_{cl} \rangle}e^{-S_{cl}}Z_{qu}.
\end{equation}
where,
\begin{equation}
\label{eq:class}
S_{cl}=\frac{1}{4\pi \alpha}\left( \int d^{2}z(\partial X_{cl}
  \bar{\partial}\bar{X}_{cl}+\bar{\partial} X_{cl} \partial \bar{X}_{cl} \right).
\end{equation}
$X_{cl}$ must satisfy the string equation of motion and possess the correct
asymptotic behaviour near the triangle vertices. This behaviour is determined
using the analogy between open strings at brane intersections and closed
strings on orbifolds, which we now discuss. 

\subsection{Boundary conditions and twist vertices}\label{subsec:twists}
\begin{figure}[t]
\begin{center}
\epsfig{file=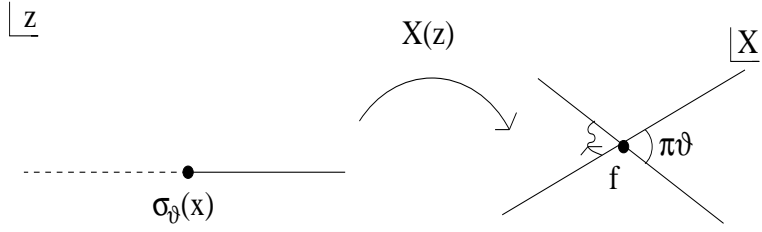, height=30mm, width=100mm} 
\caption{A twisted open string. \label{fig:twisted}}  
\end{center}
\end{figure}

Consider an open string stretched between two intersecting D-branes at an
angle $\pi \vartheta$. Solving the string equation of motion using the appropriate boundary conditions, we
obtain the mode expansion,
\begin{equation}
\label{eq:mode}
\partial X(z)=\sum_{k}\alpha_{k-\vartheta}(z-x)^{-k+\vartheta-1}.
\end{equation}
Here the  worldsheet coordinate, $z=-e^{\tau-i\sigma}$, has
domain the upper-half complex plane. This can be extended to the entire
complex plane using the `doubling trick', i.e. we define,
\begin{equation}
\partial X(z)= \left\{ \begin{array}{ll}
                         \partial X(z) & \mbox{Im}(z) \geq 0 \\
                         \bar{\partial}\bar{X}(\bar{z}) & \mbox{Im}(z) <
                         0
                        \end{array} \right.,
\end{equation}
and similarly for $\partial \bar{X}(z)$. With this extension, the mode
expansion in~(\ref{eq:mode}) is identical to that of a
closed string state in the presence of a $\mathbb{Z}_{N}$ orbifold twist
field~\cite{dixon} (with the replacement $\vartheta=\frac{1}{N}$). Therefore,
an open string stretched between two intersecting D-branes is analogous
to a twisted closed string state on an orbifold. Hence, we must introduce a
twist field $\sigma_\vartheta(x)$\footnote{Since we
are considering tree level amplitudes, all open string vertices will be
conformally mapped onto the real axis.} for the open string. Such a twist
field changes the boundary conditions of $X$ to be those required at the
intersection point, $X(x)=f$, of two D-branes. This is achieved by
introducing a branch cut in the complex plane, as depicted in~\ref{fig:twisted}. 
We can easily obtain the OPEs,
\begin{equation}
\label{eq:opes}
\begin{array}{l}
\partial X(z) \sigma_{\vartheta}(x) \sim
(z-x)^{-(1-\vartheta)}\tau_{\vartheta}(x), \\
\partial \bar{X}(z) \sigma_{\vartheta}(x) \sim
(z-x)^{-\vartheta}\tau'_{\vartheta}(x),
\end{array}
\end{equation}
where $\tau'_{\vartheta}$ and $\tau_{\vartheta}$ are excited twists. Also,
the local monodromy conditions for transportation
around $\sigma_{\vartheta}(x)$ are,
\begin{equation}
\label{eq:mono}
\begin{array}{l}
\partial X(e^{2 \pi i}(z-x))=e^{2\pi i\vartheta}\partial X(z-x), \\
\partial \bar{X}(e^{2 \pi i}(z-x))=e^{-2\pi i\vartheta }\partial \bar{X}(z-x).
\end{array}
\end{equation}
These will be employed in the next subsection, where we can now discuss the
classical solutions required for a determination of the three point amplitude.

\subsection{Classical solutions and global monodromy} \label{subsec:
  monodromy}
The three point function requires three twist vertices,
$\sigma_{\vartheta_{i}}(x_{i})$, corresponding to the three twisted string
states at the D-brane intersections. We obtain the asymptotic behaviour of
$X_{cl}$ at each of the D-brane intersections from the OPEs~(\ref{eq:opes}),
\begin{equation}
\begin{array}{ll}
\partial X(z) \sim (z-x_{i})^{-(1-\vartheta_{i})} & \mbox{as } z\rightarrow
x_{i}, \\
\partial \bar{X}(z) \sim (z-x_{i})^{-\vartheta_{i}} & \mbox{as } z\rightarrow
x_{i}.
\end{array}
\end{equation}
Then our classical solutions are determined, up to a
normalisation constant to be,
\begin{equation}
\begin{array}{ll}
\partial X_{cl}(z)= a\omega(z), & \partial \bar{X}_{cl}(z)=\bar{a}
\omega'(z), \\
\bar{\partial} X_{cl}(\bar{z})= b\bar{\omega}'(\bar{z}), & \bar{\partial} \bar{X}_{cl}(\bar{z})=\bar{b}\bar{\omega}(\bar{z}),
\end{array}
\end{equation}
where,
\begin{equation}
\begin{array}{ll}
\omega(z)=\prod_{i=1}^{3} (z-x_{i})^{-(1-\vartheta_{i})}, &
\omega'(z)=\prod_{i=1}^{3} (z-x_{i})^{-\vartheta_{i}}.
\end{array}
\end{equation}
The contribution to $S_{cl}$ coming from the antiholomorphic solution
diverges and hence we must set $b=0$.

We can now determine the remaining normalisation constant from the global
monodromy conditions, i.e. the transformation properties of $X$ as it is transported
around more than one twist operator, such that the net twist is zero. These
conditions are derived from the transformation of X around a single twist
vertex, which is,
\begin{equation}
\label{trans}
X(e^{2\pi i}z,e^{-2\pi i}\overline{z})= e^{2\pi i\vartheta}X +(1-e^{2\pi i\vartheta})f,
\end{equation}
where f is the intersection point of the two D-branes. This can be seen from
the local monodromy conditions~(\ref{eq:mono}) and the fact that $f$ must be
left invariant. The global monodromy of $X$ is then simply a product of such
actions.

\begin{figure}[t]
\centering
\epsfig{file=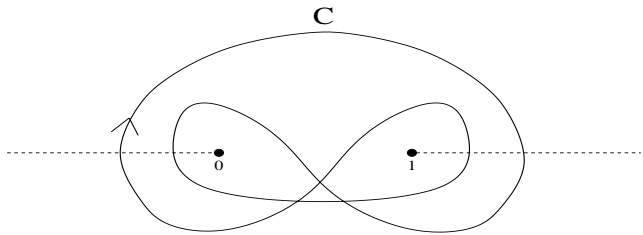, height=30mm, width=85mm}
\caption{The Pochammer loop.}
\label{fig:curve}
\end{figure}

In this three point case, there exists only a single closed loop, C, with net twist zero.
This contour is a Pochammer loop and is depicted in figure~\ref{fig:curve}.
Here we have set $x_{1}=0, x_{2}=1$ and $x_{3} \rightarrow \infty$ using
$SL(2,\mathbb{R})$ invariance and the dashed lines denote branch cuts. Evaluating the global monodromy condition,
\begin{equation} 
\label{eq:globmono}
\Delta _{C}X=\oint _{C}dz\partial X(z)+\oint
_{C}d\overline{z}\bar{\partial}X(\bar{z}),
\end{equation}
with the left hand side being determined by repeated applications
of~(\ref{trans}) and the right hand side simply by integration, we can
determine our normalisation constant $a$.

Finally, in order to fully calculate $S_{cl}$, we must
employ the methods developed in~\cite{KLT} to express the integral
over the complex plane in terms of holomorphic and anti-holomorphic
contours. We then determine the classical contribution to the three point amplitude to be,
\begin{equation}
S_{cl}=\frac{1}{2 \pi \alpha}\left( \frac{\sin
      \pi \vartheta_{1} \sin \pi \vartheta_{2}}{2 \sin \pi \vartheta_{3}}|f_{1}-f_{2}|^{2}\right).
\end{equation}
As commented earlier, this is simply the area of the triangle swept out by
the worldsheet. However, our basic methodology can now be extended to the four
point amplitude and we have illustrated the fact that the D-brane geometry
is encoded in the conformal field theory. Notice that this exponential
dependence on worldsheet area leads to a natural mechanism for hierarchy
generation as pointed out in~\cite{ibws}.

For any order amplitude, we also have contributions coming
from worldsheets which wrap the internal space. However, since the classical
contribution has the form $e^{-\frac{1}{2 \pi \alpha}\mbox{Area}}$, and the
wrapping contributions in general have far larger area then the single
unwrapped case, we have determined the leading order contribution to the amplitude. 

Finally, to fully determine the three point amplitude we also require the
quantum contribution. However, since this is independent of the generation
number, it is simply an overall factor in the case of Yukawa
interactions. Furthermore, it can be obtained from factorisation of the four
point amplitude to which we now proceed.

\section{The general four point amplitude}\label{sec:4pt}
We now outline the calculation of the general four point amplitude, this is
required when there are four indpendent sets of D6-branes. The complete
calculation can be found in~\cite{paper3}.

\subsection{Classical contribution to the four point amplitude}\label{subsec:
  class4pt}
Beginning with the classical contribution, we employ an identical methodology
to the three point case. The classical solutions we require are,
\begin{equation}
\begin{array}{ll}
\partial X_{cl}(z)= a\omega(z), & \partial \bar{X}_{cl}(z)=\bar{a}
\omega'(z), \\
\bar{\partial} X_{cl}(\bar{z})= b\bar{\omega}'(\bar{z}), & \bar{\partial} \bar{X}_{cl}(\bar{z})=\bar{b}\bar{\omega}(\bar{z}),
\end{array}
\end{equation}
where,
\begin{equation}
\begin{array}{ll}
\omega(z)=\prod_{i=1}^{4} (z-x_{i})^{-(1-\vartheta_{i})}, &
\omega'(z)=\prod_{i=1}^{4} (z-x_{i})^{-\vartheta_{i}}.
\end{array}
\end{equation}
However, the antiholomorphic solution now contributes to $S_{cl}$ and we
must determine both $a$ and $b$ from the global monodromy conditions.
To match this requirement, we now have two independent Pochammer loops to
which we apply our global monodromy condition~(\ref{eq:globmono}). This
arises as now there are four twist vertices which can be set to the positions
$0, x, 1$ and $x_{4} \rightarrow \infty$. Hence, we have a Pochammer contour
looping around $0$ and $x$ and another around $x$ and $1$. Therefore, we have
two conditions allowing us to determine our two normalisation constants.

As before, we calculate the integrals in $S_{cl}$ using the methods
of~\cite{KLT}. Then, combining with the expressions for our normalisation
constants, and  employing some algebra we obtain,
\begin{equation}
S_{cl}(x)=\frac{\sin(\pi\vartheta_{2})}{4\pi\alpha'}\left(\frac{((v_{12}\tau-v_{23})^{2}+\gamma\gamma'(v_{12}(\beta+\tau)+v_{23}(1+\alpha\tau))^{2})}{(\beta+2\tau+\alpha\tau^{2})}\right),
\end{equation}
where,
\begin{equation}
\begin{array}{llll}
\alpha=-\frac{\sin(\pi\vartheta_{1}+\pi\vartheta_{2})}{\sin(\pi\vartheta_{1})}, & \beta=-\frac{\sin(\pi\vartheta_{2}+\pi\vartheta_{3})}{\sin(\pi\vartheta_{3})},\\
\gamma\gamma'=\frac{\sin(\pi\vartheta_{1})\sin(\pi\vartheta_{3})}{\sin(\pi\vartheta_{2})\sin(\pi\vartheta_{4})},&
\tau = \left|\frac{F_{2}}{F_{1}}\right|,
\end{array}
\end{equation}
and
\begin{equation}
\begin{array}{ll}
F_{1}=\int_{0}^{x}\prod_{j=1}^{3} (y-x_{j})^{-(1-\vartheta_{j})}dy,  & F_{2}=\int_{x}^{1}\prod_{j=1}^{3} (y-x_{j})^{-(1-\vartheta_{j})}dy.
\end{array}
\end{equation}

\subsection{Quantum contribution to the four point amplitude}
\label{subsec:qu4pt}
We now briefly discuss the quantum part of the four point amplitude. We assume
four fermions localised at the intersections of four independent sets
of D-branes. The tree-level amplitude is given by a disc diagram with four
vertex operators, $V^{(a)}$, in the -1/2 picture. Using $SL(2,\mathbb{R})$
invariance, we write the ordered amplitude as,
\begin{equation}
\begin{array}{l}
(2\pi)^{4}\delta^{4}(\sum_{a}k_{a})A(1,2,3,4)= \\
\frac{-i}{g_{s}l_{s}^{4}}\int_{0}^{1}dx\langle
V^{(1)}(0,k_{1})V^{(2)}(x,k_{2})V^{(3)}(1,k_{3})V^{(4)}(\infty,k_{4})\rangle.
\end{array}
\end{equation}
The required vertex operators for the fermions are, 
\begin{equation}
V_{i}^{(a)}(x_{a},k_{a})=const\,\,\lambda^{a}u_{\alpha}^{(i)}S_{i}^{\alpha}\sigma_{\vartheta_{i}}e^{-\phi/2}e^{ik_{a}.X}(x_{a}),
\end{equation}
where $u_{\alpha}$ is the space time spinor polarization, $S^{\alpha}$
is the spin-twist operator and $e^{-\phi/2}$ is the contribution
from the superconformal ghosts. The spin-twist operator arises from
bosonization of the fermionic worldsheet fields and is of the form,
\begin{equation}
S_{i}^{\alpha}=\prod_{l=1}^{5}:\exp(iq_{i}^{l}H_{l}):
\end{equation}
 where for D6-branes intersecting at angles we have, 
\begin{equation}
q_{i}^{l}=\left(\pm\frac{1}{2},\pm\frac{1}{2},\vartheta_{i}^{1}-\frac{1}{2},\vartheta_{i}^{2}-\frac{1}{2},\vartheta_{i}^{3}-\frac{1}{2}\right).
\end{equation}
The relative sign of the first two entries determines the helicity of the
fermion, and $\vartheta_{i}^{m=1,2,3}$ are the angles of the $i'$th
intersection of the D-branes in the $m'$th complex plane.

The correlation function can be computed using OPEs and wick
contraction. However, we run into the problem of not being able to determine
the OPEs of two twist fields. We circumvent this using a method developed
in~\cite{dixon}. In outline, we determine the
correlation function of four twist vertices by first calculating,
\begin{eqnarray}
\frac{\langle
  T(z)\prod_{i=1}^{4}\sigma_{\vartheta_{i}}\rangle}{\langle\prod_{i=1}^{4}\sigma_{\vartheta_{i}}\rangle} & = & \lim_{w\rightarrow z}[g(z,w)-\frac{1}{(z-w)^{2}}].
\end{eqnarray}
Here $g(z,w)$ is the usual Green function for the closed string, whose
functional form may be determined from various asymptotics. We also know
the OPE of the stress energy tensor, $T(z)$, with a twist field. Combining
these expressions gives rise to a set of differential equations for $\langle
\prod_{i=1}^{4} \sigma_{\vartheta_{i}} \rangle$, containing a term with an
unknown constant. This may be determined from global monodromy conditions,
resulting in,
\begin{equation}
\langle\prod_{i}\sigma_{\vartheta_{i}}\rangle=|J|^{-\frac{1}{2}}x_{\infty}^{-\vartheta_{4}(1-\vartheta_{4})}x^{\frac{1}{2}(\vartheta_{1}+\vartheta_{2}-1)-\vartheta_{1}\vartheta_{2}}(1-x)^{\frac{1}{2}(\vartheta_{2}+\vartheta_{3}-1)-\vartheta_{2}\vartheta_{3}},
\end{equation}
where $|J|=|F_{1}\Vert F_{2}'|+|F_{2}\Vert F_{1}'|$ and $F'_{i}$ is obtained
from $F_{i}$ by the substitution $\vartheta_{i} \rightarrow 1-\vartheta_{i}$.

Piecing together the correlation functions of all the relevant fields and
including contributions from each torus subfactor, we obtain the full four point amplitude,
\begin{equation}
  \begin{array}{l}
    A ( 1, 2, 3, 4 )=- g_s \alpha' Tr[\lambda^1 \lambda^2 \lambda^3
    \lambda^4 + \lambda^4 \lambda^3 \lambda^2 \lambda^1 ] \\
    \int^1_0 d x \hspace{0.25em} x^{- 1 - \alpha' s} ( 1 - x )^{- 1 - \alpha'
    t}  \frac{1}{\prod_m^3 |F^{m}_{1}\bar{F'}_{2}^{m}-\bar{F'}_{1}^{m}F_{2}^{m}|^{1 / 2}}  \\
    \times \left[ \bar{u}^{( 2 )}
    \gamma_{\mu} u^{( 1 )} \overline{} \bar{u}^{( 4 )} \gamma^{\mu} u^{( 3 )}
    \right] \sum e^{- S_{c l} ( x )},
  \end{array}
\end{equation}
\normalsize
where,
\small
\begin{equation}
F^{m}_{i}=\int_{x_{i}}^{x_{i+1}}(y)^{-(1-\vartheta^{m}_{1})}(y-x)^{-(1-\vartheta^{m}_{2})}(y-1)^{-(1-\vartheta^{m}_{3})}dy.
\end{equation}
and $F'^{m}_{i}$ is again obtained from $F^{m}_{i}$ by the substitution $\vartheta_{i} \rightarrow 1-\vartheta_{i}$.

\section{Summary and conclusions}
We have outlined the techniques required to determine
interactions between states localised at D-brane intersections. This was
illustrated with an application to Yukawa interactions. We have also performed
the first calculation of the general four point amplitude. This result is
required in the case of four independent stacks of D-branes with the relevant
interaction being of the form $q_{L}q_{R} \rightarrow e_{L}e_{R}$. Our
results can also be applied to various phenomenological issues such as flavour
changing neutral currents~\cite{ams} and proton decay~\cite{witten}.

\section*{Acknowledgments}
This talk is based on work done in collaboration with S.A.Abel and the author
is supported by a PPARC studentship.

\end{document}